\documentclass[11pt]{article}
\usepackage{Blois,epsfig}
\usepackage{amssymb}
\usepackage{amsmath}

\def\be{\begin{equation}}
\def\ee{\end{equation}}
\def\bea{\begin{eqnarray}}
\def\eea{\end{eqnarray}}

\newcommand{\g}{\gamma}

\begin{document}
\vspace*{2cm}
%\begin{center}
%\Large{\textbf{XIth International Conference on\\ Elastic and Diffractive 
%Scattering\\ Ch\^{a}teau de Blois, France, May 15 - 20, 2005}}
%\end{center}

\vspace*{2cm}
\title{REMARKS ON  DIFFRACTIVE PRODUCTION OF THE HIGGS BOSON}

\author{R. PESCHANSKI}
\address{Service de Physique Th{\'e}orique, CEA/Saclay,
91191 Gif-sur-Yvette Cedex, France\\
URA 2306, unit{\'e} de recherche associ{\'e}e au CNRS}

\maketitle
\abstracts{Central diffractive production of the Higgs boson has
recently received much attention as a potentially interesting production mode 
at the LHC. We shall  review some of the wishes and realities encountered in 
this field. Theoretical open problems of diffractive dynamics are  involved 
in  making accurate predictions for the LHC, among which the most crucial is 
understanding  factorization breaking in hard diffraction.}

\section{Original concept}
\begin{figure}[hb]
\begin{center}
\epsfig{file=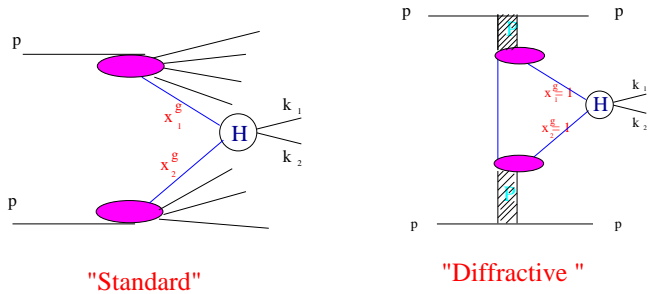,width=8.5cm}
\caption{``Standard'' vs. ``Diffractive'' Higgs boson production.}
\end{center}
\label{F1}
\end{figure}

The potential interest of central diffractive production of the Higgs boson 
is  illustrated in Fig.1; while standard production via gluon-gluon fusion 
can reach high cross-sections, the study of the Higgs boson will be  
uneasy due to  accompanying particles and backgrounds, especially if it takes place 
in the 
low-mass range where $\g\g$ is the main observational decay mode. The original 
guiding line for central diffractive production   \cite{Bialas:1991wj} is to 
compensate the weak cross-sections by a cleaner signal, and precise production 
kinematics   \cite{Albrow:2000na} thanks to the  tagging of the diffracted 
protons.
The basic diagrams corresponding to diffractive Higgs boson production 
correspond to all combinations of double gluon exchanges in various  
 ways \cite{Bialas:1991wj}. Their main property, which remains valid even if the 
gluon propagators are assumed non-perturbative, is the resummation 
property  pictured in Fig.2; the sum of diagram contributions boils down to the 
on-shell 
convolution of (non-perturbative) gluon exchange contribution times the simplest  
(non-perturbative) $gg \to H$ fusion diagram contribution.
\begin{figure}[hb]
\begin{center}
\epsfig{file=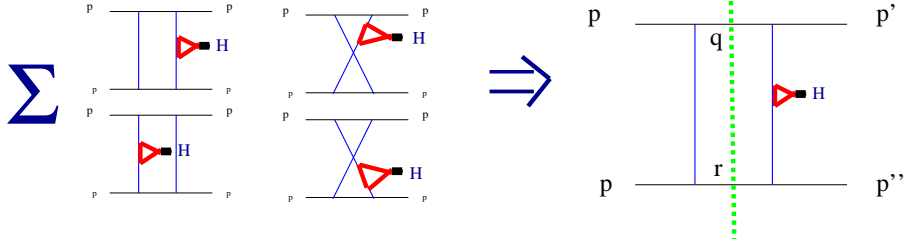,width=12cm}
\caption{Higgs boson production: Factorization of lowest-order diagrams.}
\end{center}
\label{F2}
\end{figure} 

\section{Wishes and realities}

The original dynamics were presented in the framework of the simplest 
lowest-order diagrams. They have 
been considered  in the framework of non-perturbative gluon propagators with a 
fixed  $\cal O$(1) coupling constant. 

What are the problems we are facing when considering ``higher order'' 
contributions? They are responsible for the production of ``extra'' particles 
which may accompany  the Higgs boson production. It is not clear whether these 
extra particles are taken into account by the calculation of  the 
original paper  \cite{Bialas:1991wj}. They may change drastically the estimate of 
Higgs production 
if one insists on putting a veto on the production of these  extra particles. 
In the case extra particles are produced in the central region of rapidity, one 
speaks of ``inclusive'' diffractive production of the 
Higgs boson, while the term ``exclusive'' has been kept for the  case where a strict 
veto on extra particles has been imposed. 
They are both sketched in Fig.3.
\begin{figure}[hb]
\begin{center}
\epsfig{file=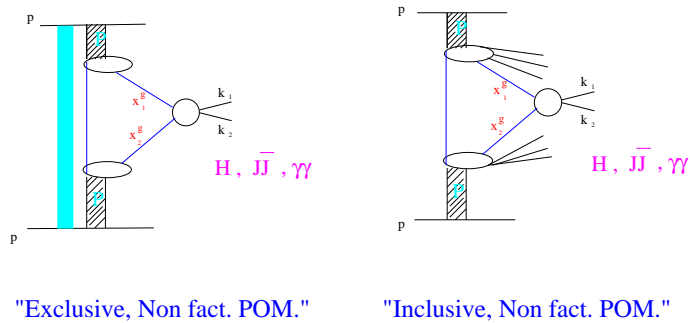,width=9cm}
\caption{High mass states production: exclusive vs. inclusive. Left: The 
``Pomeron-induced'' exclusive interaction is corrected for initial state 
radiation. Right: The ``Pomeron-induced'' inclusive interaction produces soft 
particles together with the heavy mass state: Higgs boson, dijet, diphoton.}
\end{center} 
\label{F3}
\end{figure} 
On the left of Fig.3 is shown the ``Pomeron-induced''  model, which comes from 
a modification of the Bialas-Landshoff  model. For exclusive production  
 \cite{ourpap}  the original combination of diagrams is modified by 
taking into account the veto on particle radiation due to the initial states 
or ``rapidity gap survival''. On the right,  one sees a typical ``inclusive'' 
production contribution as originally proposed in Ref.  
 \cite{Boonekamp:2001vk}. In this model, the extra particles can be considered as 
``Pomeron remnants'' in a Pomeron-Pomeron collision.

As for the models, the Pomeron-induced models inspired by the original 
Bialas-Landshoff approach make use of the soft Pomeron interaction to describe 
the diffractive coupling  to the incident particles. Two other competitive 
approaches make a different ansatz.  In the ``proton-induced'' 
model  discussed in Ref.  \cite{durham}, the gluon exchange + fusion mechanism is 
considered for purely exclusive processes, and   is expected to allow for  a 
perturbative treatment. Besides 
the  ``rapidity gap survival'' (RGS) soft correction factor, the cross-section 
gets corrected 
by the perturbative Sudakov form factors. Their effect is   severely cutting the 
production of 
high mass states. On the other hand, the energy dependence, being driven by the 
perturbative QCD rise of 
the gluon distribution in the proton, is quite rapid.  In a third approach, 
the Soft Color Interaction (SCI) model   \cite{Enberg:2002id}, the colored $gg$ 
fusion contribution is corrected by assuming a long distance neutralization 
of color.

The problem of exclusive production is that, up to know, it belongs to  
physicists' wishes but  not yet belongs to physics realities. Indeed, the 
idea of testing central diffractive production in the perspective of the Higgs 
boson production is to look for known color singlet massive states such as 
dijets or diphoton of large masses, as sketched in Fig.3. Inclusive production 
of massive dijets has been tested (hard diffractive production of jets has already 
been observed by the UA8 Collaboration at CERN and  at HERA) at the Tevatron Run I  
 \cite{Affolder:2000hd}, and already copiously produced till the beginning of 
Run II. In the same time, there is yet no evidence (and thus only upper 
bounds) for  exclusive production   \cite{Gallinaro:2005uh}. Hence models for 
exclusive production cannot ``calibrate'' their cross-sections on real events. 
Even if models of Refs.   \cite{ourpap} and   \cite{durham} give similar 
cross-section predictions for a low mass standard Higgs boson at the LHC (of 
the order of a few femtobarns) they greatly differ when the mass or the energy 
varies. On contrary the models for inclusive production, such as Refs.  
 \cite{Boonekamp:2001vk} and \cite{Enberg:2002id}, are better validated thanks to 
their description of the observed dijet production. 
It is clear that the selection of viable models  will be clarified by  
hard diffraction studies  at the Tevatron. 
\begin{figure}[hbt] 
\begin{center} 
\includegraphics[width=6cm]{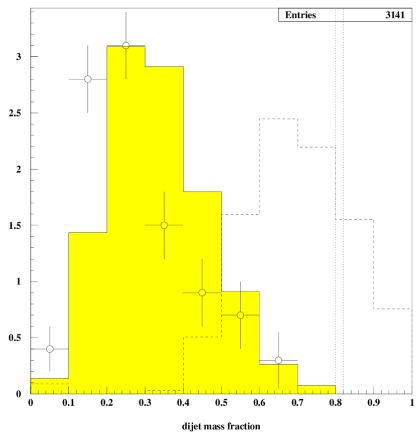}
\caption{Dijet Mass fraction at the Tevatron. Data are from Ref.[7]; shaded histogram: 
model predictions from Ref.[4]; dashed  
(resp. dotted) contour: simulation of exclusive 
production at detector (resp. parton) level.}
\end{center}
\label{F4} 
\end{figure}

\section{The ``Sesame'' of diffraction: understanding factorization breaking}

The key question for the efficiency of hard diffractive production, in 
practice for the evaluation of cross-sections, is factorization breaking. 
It is well-known that the ratio of diffractive over non-diffractive hard 
events at the Tevatron compared to HERA is a factor circa 1/10 for similar 
kinematical variables. This factor  came as a surprise, but after all,  the 
soft interaction between the incident hadrons, absent at HERA (except for 
e.g.  photoproduction, which is an interesting 
problem), is expected to interfere with the hard interaction producing the 
heavy state, and thus to break factorization. The whole question is to find which 
is 
the mechanism of this interference between soft and hard processes both 
present in the same collision. Here we only rely on phenomenological  models 
since little is theoretically known  about soft interactions, non-perturbative 
in terms of QCD.
\begin{figure}[hbt] 
\begin{center} 
\includegraphics[width=8cm]{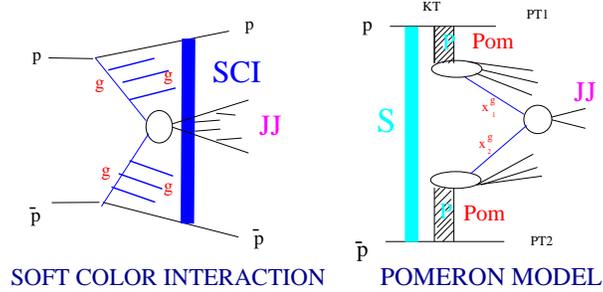}
\caption{Models of factorization breaking.
{Left:} ``soft color interaction'' ;
{Right:}  ``rapidity gap survival'')}
\end{center}
\label{F5} 
\end{figure}
In order to illustrate the problem, we consider  two popular approaches of 
factorization breaking mechanisms, namely the one used in the SCI approach 
 \cite{Enberg:2002id} and the  
RGS approaches used  in (exclusive) Pomeron  \cite{ourpap} and proton  
 \cite{durham} induced models. In the former (SCI), a soft color interaction  
correcting the standard colored exchange  may form a final singlet state 
allowing for diffraction. This  is indeed a {\it gap-creating} mechanism. By 
contrast, in the  RGS approaches, the initial hard, factorized, diffractive 
mechanism forming gaps is hidden by the interaction between initial hadrons. 
It is thus a {\it gap-destroying} mechanism. 
\begin{figure}[hb] 
\begin{center} 
%\hspace{0.5cm}
%\begin{minipage}[t]{55mm}
\includegraphics[width=7cm]{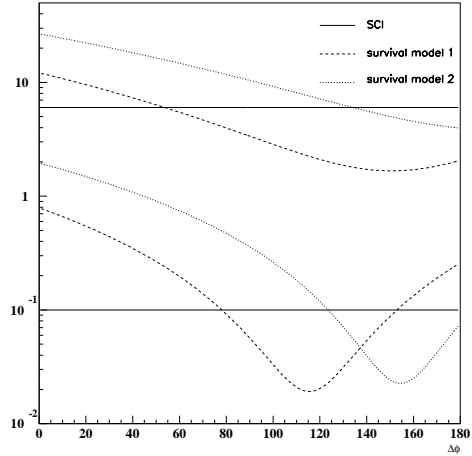}
%\end{minipage}
%\hspace{\fill}
%\begin{minipage}[t]{75mm}
%\includegraphics[width=6.8cm]{fig1b.eps}
%\end{minipage}
%\hspace{0.5cm}
\caption{Shape in $\Delta \Phi$ between the outgoing $p$ and $\bar{p}$ for
SCI and RGS models. The upper plot is for asymmetrical cuts in $P_T^2$ 
($|P_T^2(p)| > 0.6$, $|P_T^2(\bar{p}p)| > 0.1$ GeV$^2$) and the lower ones
for symetric cuts
on $P_T^2$ ($|P_T^2(p)| > 0.5$, $|P_T^2(\bar{p}p)| > 0.5$ GeV$^2$). 
We note the different behaviour for SCI and  RGS models where the
minimum is close to $\Delta \Phi \sim$ 180 degrees (resp. 130 degrees) for
asymmetrical (resp.symmetric) cuts in $P_T^2$. The  $\Delta \Phi$ dependence can be 
studied using the FPD detector, see next Fig.7.}
\end{center}
\label{F6}
\end{figure}

Interestingly enough, these two models lead to very different predictions for 
central diffractive Higgs boson production at the LHC for a Higgs of mass 
around 120 GeV. While RGS models predict an overall small but sizeable 
cross-section of the order of the femtobarn, the  SCI model predicts a 
completely negligible double-diffractive cross-section.

It is thus important to find ways of discrimination between models. For this sake,
the experimental diffraction physics programme at the Tevatron will be crucial. As 
an illustration, let us consider  \cite{Kupco:2004fw} the inclusive diffractive 
production of dijets and suppose that we measure the full transverse momenta of 
the outgoing forward proton and antiproton, namely $P_T$ and the azimuthal angle 
$\Delta \Phi$ between the two particles. It appears, see Fig.6,  a significant 
difference 
between the RGS and SCI models. In Fig.6, we also compare two different RGS 
models, we call them model 1  \cite{model1} and model 2  \cite{bialas}, differing 
only through the inclusion or not of soft inelastic diffractive channels in the 
calculation (see   \cite{Kupco:2004fw} for details).

The difference between  RGS and SCI models can be easily traced back to the basic 
contrast between a {\it gap-destroying} and a {\it gap-creating} mechanism. In the 
former case, the correction is expressed as a destructive interference between two 
terms, namely
\begin{equation}
{\cal A}(p_{T1},p_{T2}, \Delta \Phi) =
 {\cal A}^{h} -{\cal A}_{SP}\ {\bf *}{\cal A}^{h} \ \equiv \int d^2{\bf 
k}_T\ \left\{ 1 -{\cal A}_{SP} \right\}({\bf k}_T) \ {\cal A}^{h}({\bf 
p}_{T1}\!-\!{\bf k}_T,
{\bf p}_{T2}\!+
\!{\bf k}_T) 
\ ,  
\nonumber
\end{equation}
where ${\cal A}_{SP}$ is the ${\cal O}(1)$ ``effective'' (including inelastic 
diffraction, model 1) or genuine (model 2) elastic $p\bar p$ amplitude. ${\cal 
A}^{h}$ is the uncorrected hard diffractive amplitude. On the contrary, SCI 
models, at least in their present version (some modification of the prediction could 
come from a different  simulation of hadronization), do not lead to striking 
structures 
since, schematically,  they just weight the contributions of standard non 
diffractive diagrams to cross-sections by an universal probability factor. No 
strong interference effect is expected in this case.

An experimental  measurement is not unrealistic, even if delicate,  as azimuthal 
asymmetries could be  checked at the FPD detectors included in the D0 apparatus, 
see Fig.7. 
\begin{figure}[hbt] 
\begin{center} 
\includegraphics[width=8.5cm]{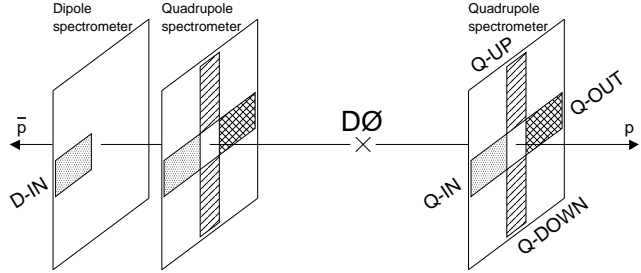}
\caption{Schematic view of the FPD detector.  This setting may allow for a study of 
azimuthal dependence of the outgoing hadrons.}
\end{center}
\label{F8} 
\end{figure}

\section{Tasks in diffraction theory}
At the present stage of our knowledge and since our experimentalists friends are 
preparing  for the opening of the LHC, it is worth trying to select the  
valuable theoretical questions raised by the diffractive production channel of the 
Higgs boson.

\begin{itemize}

\item {\it Evaluation of the inclusive production cross-section at the LHC.} The 
formation of hard diffractive dijets in the central region of the Tevatron is now 
certain. The precise description of their observed characteristics and a motivated 
 prediction of the energy dependence towards the LHC conditions seem possible in 
the next period.

\item {\it Evaluation of QCD corrections.} The radiative QCD corrections are 
expected to be strong, under the form of Sudakov form factors in the 
``proton-induced'' model of exclusive production. On the other hand, QCD corrections 
are also  present in 
the tail of inclusive production, i.e. the end of the dijet mass spectrum, see 
Fig.4, of the inclusive 
production of dijets at Tevatron. I suggest a thorough comparison between the two  
approaches, and more generally on the exclusive approaches compared to the 
``quasi-exclusive'' ones dealing with the tail of the inclusive distribution. 
In fact, in practice, the experimentalists will find difficult to give a criterium 
selecting strictly exclusive production.

\item {\it Evaluation of ``soft'' corrections.} We have seen that there exist ways 
to disentangle rapidity-gap creating from rapidity-gap destroying types of models.
Identifying the non-perturbative source of factorization breaking is a major task 
for the theoreticians interested into diffractive mechanisms. It has a major impact 
on the evaluation of cross-sections, since at the LHC, these effects are predicted 
to hide a large fraction of the interesting events (another more experimental 
problem, but also related with soft hadronic physics, is the piling-up due to  some 
or many minimum-bias collisions, depending on the working luminosity).

\end{itemize}
The diffractive production could be an interesting complementary device for Higgs 
boson search at the LHC.  Moreover, if diffractive production of known 
heavy states is confirmed, then it is worth investigating SUSY particle production 
 \cite {durham,Boonekamp:2005yi}. It however requires some theoretical work to 
obtain a 
reasonable evaluation of the expected cross-sections. Fortunately enough, the 
diffractive program at the Tevatron and the search for known massive states which 
can be  diffractively produced: dijets, diphotons, $t\bar t$, WW (through QED 
production, see e.g.  \cite{Boonekamp:2005yi}) will give quite a few instructive 
answers in the near and next-to-near future.

\vspace{-.3cm}

\section*{Acknowledgements}
I want to thank the organizers of the Blois conference for the invitation and my 
collaborators M.Boonekamp, J.Cammin, A.Kupco, C.Royon,  for the pleasure  of 
investigating 
new particle physics with  diffractive tools.

\end{document}